# Jointly They Edit: Examining the Impact of Community Identification on Political Interaction in Wikipedia

Jessica G. Neff[1], David Laniado[2], Karolin Eva Kappler[2], Yana Volkovich[2], Pablo Aragón[2] & Andreas Kaltenbrunner[2]

University of Southern California[1] Barcelona Media Foundation[2]

Abstract

In their 2005 study, Adamic and Glance coined the memorable phrase 'divided they blog', referring to a trend of cyberbalkanization in the political blogosphere, with liberal and conservative blogs tending to link to other blogs with a similar political slant, and not to one another. As political discussion and activity increasingly moves online, the power of framing political discourses is shifting from mass media to social media.

Continued examination of political interactions online is critical, and we extend this line of research by examining the activities of political users within the Wikipedia community. First, we examined how users in Wikipedia choose to display (or not to display) their political affiliation. Next, we more closely examined the patterns of cross-party interaction and community participation among those users proclaiming a political affiliation.

In contrast to previous analyses of other social media, we did not find strong trends indicating a preference to interact with members of the same political party within the Wikipedia community.

Our results indicate that users who proclaim their political affiliation within the community tend to proclaim their identity as a 'Wikipedian' even more loudly. It seems that the shared identity of 'being Wikipedian' may be strong enough to triumph over other potentially divisive facets of personal identity, such as political affiliation.

## Introduction

Online media have become an increasingly important source of political information in recent years. This trend emerged most notably in the 2004 U.S. presidential campaign. For the first time, political blogs served as a prominent information source regarding the campaign and candidates (Adamic & Glance, 2005), and candidates themselves began to leverage the power of online platforms to organize and raise money (CampaignsOnline.org, 2004). The trend of utilizing online platforms for political purposes, and in particular for the dissemination of information, has continued to grow in recent years. People consult political blogs, gain information about politicians, legislation and emerging social movements through their Facebook pages, and turn to online resources such as Wikipedia for up-to-date information on political issues (Smith, 2011). Given the increasing prominence of the Web, and social sites in particular, as sources of political information, it is crucial to take a closer look at the patterns of interaction and discourse that members of different political parties have around information online, because they may have important consequences for the accuracy and neutrality of political information provided online.

## Political Interaction Online

Much of the research examining political interaction online has provided support for a trend of polarization. One of the seminal studies in this area was Adamic and Glance's (2005) examination of the political blogosphere. They examined linking behavior among political blogs in the months leading up to the 2004 U.S Presidential election, and found that conservative and liberal political blogs primarily link to other blogs with their same political orientation and exhibit far fewer links to blogs that do not fall within their own political community. Using a different dataset and a different methodology, Ackland (2005) also replicated Adamic and Glance's (2005), providing additional evidence for polarization in the political blogosphere. In a similar study Hargittai, Gallo, and Kane (2007) provided further support for this trend. Their examination of linkages among political blogs also revealed a tendency for blogs to link with blogs that are ideologically similar; however, they also found evidence for cross-ideological linkages. Qualitative analysis of these linkages indicated that the vast majority of these links are used in the context of "straw-man" arguments, and therefore are not indicative of true cross-party dialogue. Blog readership also follows similar patterns of fragmentation. For example, Lawrence, Sides & Farrell (2010) found that people tend to read blogs that reinforce, rather than challenge, their political beliefs. Taken together, these studies provide strong support for the trend towards fragmentation and polarization on the political blogosphere across party lines.

The trend for polarization has also been observed in other online contexts. For example, Feller, Kuhnert, Sprenger, and Welpe (2011) considered patterns of interaction among political users on Twitter. They analyzed a sample of 2,500 German Twitter users. From this large sample they generated a subsample of 759 political users. In line with Adamic & Glance (2005), they also found patterns of preferential interaction based on political party. Users were more likely to be connected to other users who shared the same party affiliation. When linking was seen across parties, it was more frequent among parties that were more ideologically similar. Research has revealed similar findings for political Twitter users in the U.S.. Conover et al. (2011) analyzed tweets containing politically valenced hashtags in the six weeks leading up to the 2010 midterm Congressional elections. They found strong evidence of political polarization in the network of retweets, with users more likely to retweet users with whom they share the same political ideology. Similar results have been found in multiple party environments, such as the Twitter-campaign of the Spanish national elections in 2011 (Aragón et al., 2012).

In contrast to the findings from the blogosphere and Twitter, research on interactions in political newsgroups does not paint a clear picture of polarized interactions. Kelly, Fisher, and Smith (2006) analyzed the discussion networks of members of political newsgroups on Usenet and found a great deal of cross-party interaction, indicating that these newsgroups were spaces for "debate and deliberation", as opposed to "ideological echo chambers" (p. 1). They also identified distinct types of users. The vast majority of users were what they termed 'fighters', that is, they exhibited a preference to interact with members of opposing political parties and were less likely to interact with members of their own political party. They also identified a class of users they call 'friendlies', who only engage with allies (or same party users), and ignore users with opposing party affiliation and viewpoints. However, these users were much less common than the fighters. In another study of patterns of interaction in online discussion groups, Wojcieszak and Mutz (2009) also show support for cross-party engagement in online discussion groups.

The aforementioned studies have provided an unclear and somewhat conflictive picture of what cross-party political interaction looks like online. In some contexts (e.g. the blogosphere, Twitter) interactions that cross ideological divides are rare. However, in other settings (e.g. online discussion boards), there is evidence for higher rates of interactions across party lines. Taken together these finding indicate that the degree of interaction and engagement with politically dissimilar others varies across contexts.

Understanding when and why people engage in political debate and discussion online is important. The degree of interaction or insularity of political groups producing political information online can have important consequences for information consumers, because it may influence the extent to which issues are presented in a biased or neutral way. The present research seeks to address this issue, and to shed light on patterns of political interaction within the Wikipedia community.

## Political Interaction in Wikipedia

Wikipedia, the 6th most trafficked site in the world (Alexa, 2012), is arguably one of the most important information sources on the Web. In January 2012 it received 482 million global unique visitors (Wikimedia Highlights, 2012). For many people, Wikipedia is the first site they visit when they want to familiarize them with a new topic. A recent poll revealed that as of May 2010, 53% of U.S. users of the Internet sought out information in Wikipedia. A web search yields a link to a Wikipedia entry among the top three search results almost 90% of the time (Silverwood, 2012). Past research has revealed that Wikipedia entries on topics from a variety of different disciplines (Giles, 2005), including politics (Brown, 2011) are extremely accurate.

However, Wikipedia is unique when compared to other online references. In the world of online information there is professional content, some of which aims for a neutral stance and some of which has a self-proclaimed bias, and there is user generated content (UGC), which, at its core, reflects the beliefs and ideologies of those who create it. Wikipedia is built entirely on UGC; however at the same time explicitly espouses neutrality. One of the fundamental rules of the community is that all articles must be edited from a neutral point of view (NPOV). In Wikipedia, neutrality “means carefully and critically analyzing a variety of reliable sources and then attempting to convey to the reader the information contained in them clearly and accurately. Wikipedia aims to “describe disputes, but not engage in them" (NPOV, 2012). This marks a difference with respect to other communities, such as Conservapedia, created in opposition to Wikipedia to explicitly carry a conservative point of view, and self-described as "a conservative, family-friendly Wiki encyclopedia" (Conservapedia, 2012).

While neutrality is a fundamental principle of Wikipedia, members have a diverse array of beliefs and values. Therefore, it is particularly interesting to examine how diverse, and at times contentious, groups interact on the site. How is it that these people come together to create neutral content? Is there fragmentation, as we see in the blogosphere and on Twitter, or is there interaction and debate like we see in communities such as the Usenet?

Extant research has yet to consider interactions among members of different political parties on Wikipedia. However, one exception is a recent study that examined edits made to the Republican presidential candidate Mitt Romney's Wikipedia page (Fitzpatrick, 2012). Some of the most edited topics on the page were those related to controversies surrounding Romney, which have frequently been invoked in partisan debates. Findings also indicated that the peak in the number of edits made to the page coincided with the Florida Presidential primaries. As a potential explanation for the timing of this peak, the researchers suggest that perhaps users are making edits in an effort to influence public opinion. This study provides some indication that there may be partisan conflicts taking place among Wikipedia users. However, the study only looked at editing behavior in general and did not examine the political affiliations of individual users. In the present research we seek to provide a more in depth look at politics in Wikipedia by examining patterns of interaction among self proclaimed Republican and Democratic users through the lens of social identity theory.

## Social Identity Theory

Social identity theory (Tajfel & Turner, 1986) provides a theoretical framework for understanding patterns of cross-party interaction. This theory and the related self-categorization theory (Hogg, 2001; Turner, Hogg, Oakes, Reicher, & Wetherell, 1987) address how identification and categorization influence intergroup interactions. The central thrust of these theories pertains to the existence of multiple, socially defined 'selves.' They maintain that we do not have a single, static self, but rather that we have a variety of different self categorizations that may become salient depending on what context we are in (Turner et al., 1987). These categorizations may be either personal identities or social identities. A person can have any number of personal and social identities. Spears and Lea (1994) provide the following description of the self-categories available to individuals:

> …self categories can be ordered in terms of a hierarchy of abstraction and include personal identities (which distinguish the person from other individuals or in-group members) and social identities (which define them as similar to other in-group members and different from out-groups on relevant dimensions). In sum, the salient self category is highly flexible and context dependent. (p.441)

Social identity can be derived from membership in a formal group (e.g. a soccer team), but can also be derived from more abstract groups or categorizations (e.g. race, gender). Tajfel and Turner (1986) provide a broad-based description of groups, defining a group as "a collection of individuals who perceive themselves to be members of the same social category, share some emotional involvements in this common definition of themselves, and achieve some degree of social consensus about the evaluation of their group and of their membership in it" (p.15). Social identification results in a sort of "us" versus "them" dynamic, with individuals treating in-group members preferentially and discriminating against out-group members.

**Social Identity and Party Affiliation**

Social identity theory has been applied to the domain of politics, and research has demonstrated that people can develop social identities stemming from political party affiliation (Deaux, Reed, Mizrahi, & Ethier, 1995). Political identity has been offered as a theoretical explanation for the strong partisan tensions that emerge, for example, between the U.S. Republican and Democratic parties. Identification with a political party can lead an individual to selectively attend to information that supports his or her own party, while ignoring information that supports the other party (Greene, 1999). In one of the first studies of social identity in the context of U.S. politics, Greene (1999) found that the strength of an individual's party identification was a significant predictor of ratings of in-party and out-party members. Individuals with strong party identification had more favorable ratings of in-party members and less favorable ratings of out-party members, in contrast to individuals who exhibited weaker party affiliation. A later study extended these findings (Greene, 2004), and revealed that strength of party identification is also significantly related to likelihood of engaging in partisan activities (e.g. making financial contribution to a campaign, attending a campaign rally, etc.) and voting for the party in elections. Fowler and Kam (2007) also found that strength of political identification is linked to political participation. Taken together, these findings provide strong support for the claim that social identity can be derived from political party membership, and that such identification can have an important impact on perceptions of, and interactions with, members of other political parties. This insight helps us to make sense of findings from previous research on political interaction online. Individuals with strong party affiliations (e.g. political bloggers, activists who tweet) will likely prefer to interact with members of their same party, and will view same party members in a more positive light.

**Social Identity in Online Communities**

Membership in an online community may also be a source of social identity. Recent theoretical work by Ren, Kraut, and Kiesler (2007) has explored this phenomenon in greater depth. Ren et al. argue that individuals can develop attachments to online communities based on a common identity (an attachment to the community at large) and common bonds (an attachment to individual community members) (p.378). Attachments based on common identity are most relevant for the present discussion. The authors note that, “in general, common identity in the online context implies that members feel a commitment to the online community’s purpose or topic” (p. 381).

One source of common identity in online communities is task interdependence. When community members are working together to accomplish a joint task, this can foster a sense of shared identity (Ren, Kraut, & Kiesler, 2007). Wikipedia is an example of such a community. A diverse group of people comes together to create a shared good – a collaboratively authored encyclopedia. Another source of common identity is sense of community, a concept that was originally proposed in the context of offline communities (McMillan & Chavis, 1986; McMillan, 1996), which has since been extended to the virtual domain (Blanchard & Markus, 2004). Sense of community is “a feeling that members have of belonging, a feeling that members matter to one another and to the group, and a shared faith the members' needs will be met through their commitment to be together" (McMillan & Chavis, 1986, p.4). The sense of community that users feel in Wikipedia may also drive users' identification with the community. Rafaeli and Ariel (2008) have posited that the sense of community that users derive from Wikipedia may be one of their primary motivations for participation.

A study conducted by Bryant, Forte and Bruckman (2005) provides evidence for the presence of a sense of a community within Wikipedia. They described the process by which newcomers move from the periphery of the community to taking on more active and central roles. As users become more involved in the site, a transformation takes place in their identity, as they come to view themselves as *members of the tribe* and gain awareness of social roles in the community. This transformation is accompanied by a shift in activity, as members move "from a local focus on individual articles to a concern for the quality of the Wikipedia content as a whole and the health of the community" (Bryant et al., 2005, p. 9).

**Research questions**

The preceding discussion has reviewed the formation of social identity, and the influence that identity can have on intergroup dynamics. Individuals may have multiple social identities that become more or less salient depending on the social context. Members of the Wikipedia community who publicly declare their political party affiliation represent a minority of users. However, the fact that these users choose to call attention to this aspect of their identity is noteworthy. Therefore, it seems possible that either the social identity of party affiliation or of being Wikipedian could be activated in the context of this community. In the present research we examine user practices of representation and identity and examine patterns of cross-party interaction. In light of the preceding review, we pose the following research questions:

RQ1: What are the identity and representation practices of users who claim their affiliation to a party within the Wikipedia community?

RQ2: Do we see division in patterns of participation along party lines?

RQ3: Do users exhibit a preference for interacting with members of their same political party?

RQ4: Do we see differences in the emotional expression of users from different political parties?

RQ5: Does political affiliation of users affect the amount of conflict in discussions?

# Methods

## Overview

We conducted a mixed-methods analysis of patterns of activity, interaction, and identity representation practices among 1390 members of the Wikipedia community who explicitly proclaim their political affiliation as either a Republican or Democrat. In order to determine user political affiliation, user pages were examined. Content analysis was used to evaluate user representation practices in user profiles, and to categorize and thematically group the most edited articles by members of each party. Social network analysis was used to explore the research questions regarding patterns of interactions. Sentiment analysis was used to analyze the tone of the discussions in an effort to more deeply explore potential differences in the behaviors of Democrats and Republicans. Finally, content analysis was used to examine differences to conflict across and within parties, and to evaluate user representation practices in user profiles.

**User selection**

On personal user pages, users have the ability to display userboxes. A userbox is "a small colored box designed to appear only on a Wikipedian's user page as a communicative notice about the user, in order to directly (or indirectly) help Wikipedians collaborate more effectively on articles." (Wikipedia, 2012). Userboxes are customizable, and a user can choose to include any information she would like. See Figure 1 for an example of userboxes. In order to select Republican and Democratic users, userboxes that identified a user as a Republican or Democrat were manually identified and then collected automatically. Users were included as members of a party if they had a box on their user page that expressed explicit support for a party (e.g. "This user supports the GOP", "This user supports the Democratic Party") and/or a userbox that expressed support for a particular political candidate (e.g. "This user supports Barack Obama for President" or "This user supports John McCain"). In order to be able to identify different kinds of userboxes and templates, we searched in the "User" namespace for links to the articles of the Democrat and Republican party and of their major leaders. Additionally, we searched for specific sentence patterns in namespace "User", such as "This user supports the * party". Using this method to identify users resulted in a sample of 863 Democrats and 527

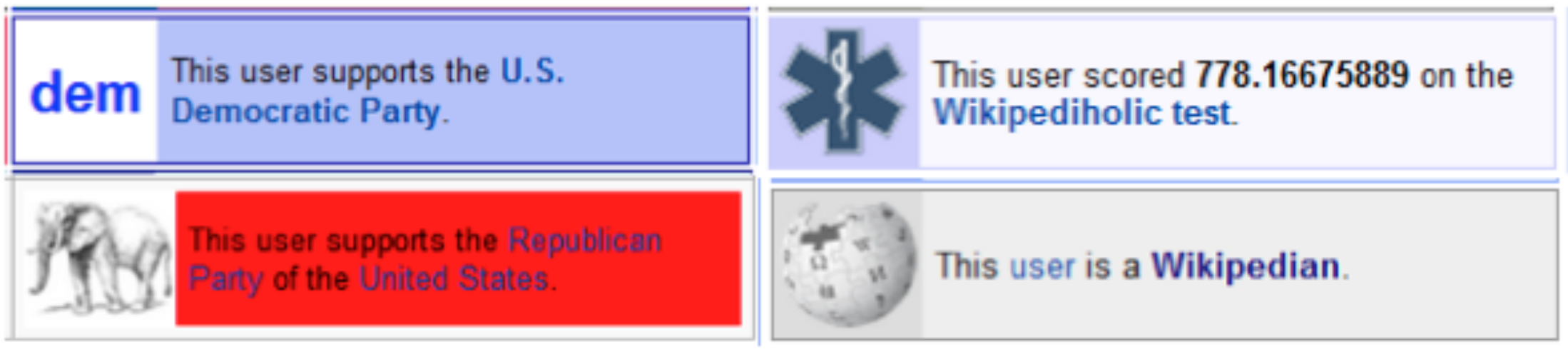

**Figure 1:** Examples of Userboxes.

Republicans, which should correspond to nearly all users who disclosed their support for one of the two major U.S. political parties in a userbox.

**Identity analysis**

Members of the Wikipedia community have the option of creating a customized user page. Pages can be personalized to reflect the preferences and interests of the individual users, and one of the primary ways that users personalize their pages is through the use of userboxes, which were described above. A qualitative analysis of the userboxes of a randomly selected sample of fifty Democratic and fifty Republican users was conducted.

First, the overall number of userboxes for each user was tallied. Next, the number of political party boxes that a user listed on his or her page was tallied. A box was counted as a party box if it explicitly expressed support for, or membership in, the Republican or Democratic Party. The number of politically oriented userboxes that were *not* political party boxes was also tallied. Politically oriented userboxes were coded as either "conservative", "liberal", or "other." Boxes coded as conservative were those that expressed what is generally considered a conservative ideology. Examples include, "This user is pro-life", "This user supports LEGAL immigration", and "this user thinks the global warming issue has been immensely exaggerated." Boxes coded as liberal were those that expressed what is generally considered a liberal ideology. Examples include, "This user supports the legalization of same-sex marriage", "This user is pro-choice", and "This user supports immigration and the right to travel freely upon the planet we share." Issues coded as "other" were those that dealt with some political issue, but are not generally assigned to a particular political ideology. Examples include, "This user wants ZERO net carbon emission from human activity", "This user is against monarchy", and "This user condemns and opposes Srebrenica Genocide denial."

Content analysis of user walls was performed on a randomly selected subsample of 100 user pages (50 Republicans, 50 Democrats). Intercoder reliability was assessed using Holsti's (1969) reliability score, which measures the percent agreement between two coders ratings. The obtained coefficient of .84 was acceptable.

**Data extraction**

Activity and interaction data came from a complete dump of the English Wikipedia, dated March 12$^{th}$ 2010. First, we considered edit activity. We counted all edits made by users in our sample to each Wikipedia article. Figure 2 shows the cumulative distribution of the number of edits per user, broken by party: the distribution for Democrats is depicted in blue, for Republicans in red. The two curves are very similar, with about 75% of the users having more than 50 comments, and about 25% of users with over 1000 edits. The major difference is that the most active users in our sample, reaching the order of 100 thousand edits, are Democrats.

We also extracted comments written by users in talk pages, i.e. special wiki pages devoted to communication among editors. We considered both *article talk pages*, where users can discuss issues concerning the corresponding articles, and *user talk pages*, or *user walls*, which are used by editors to exchange personal messages. Data were obtained by parsing the source text of talk pages and identifying user signatures and comment indentation to reconstruct the thread structure, as described in Laniado, Tasso, Volkovich and Kaltenbrunner (2011). The distribution of the number of comments per user in article talk pages (Figure 3) shows that within both parties around 50% of users wrote more than 5

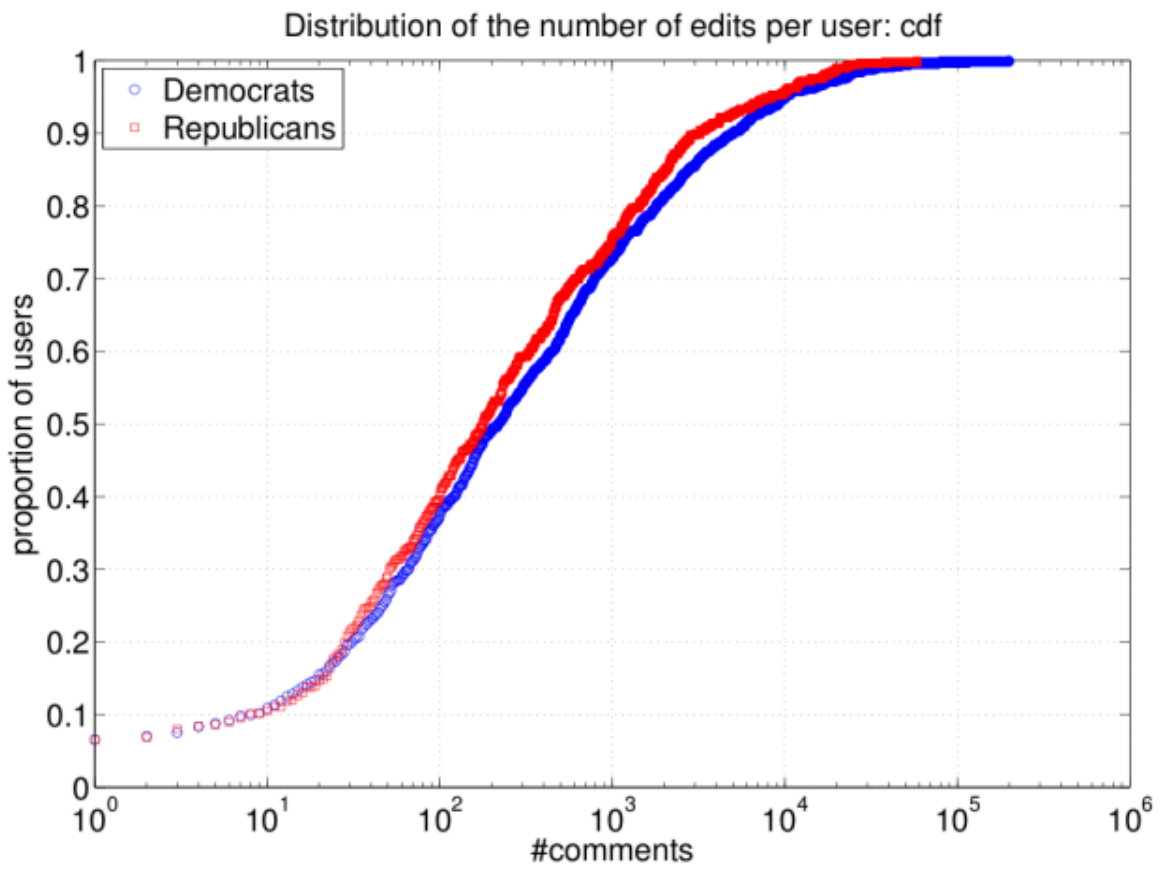

**Figure 2**: Cumulative distribution of the number of edits per user, for Democrats (blue) and Republicans (red).

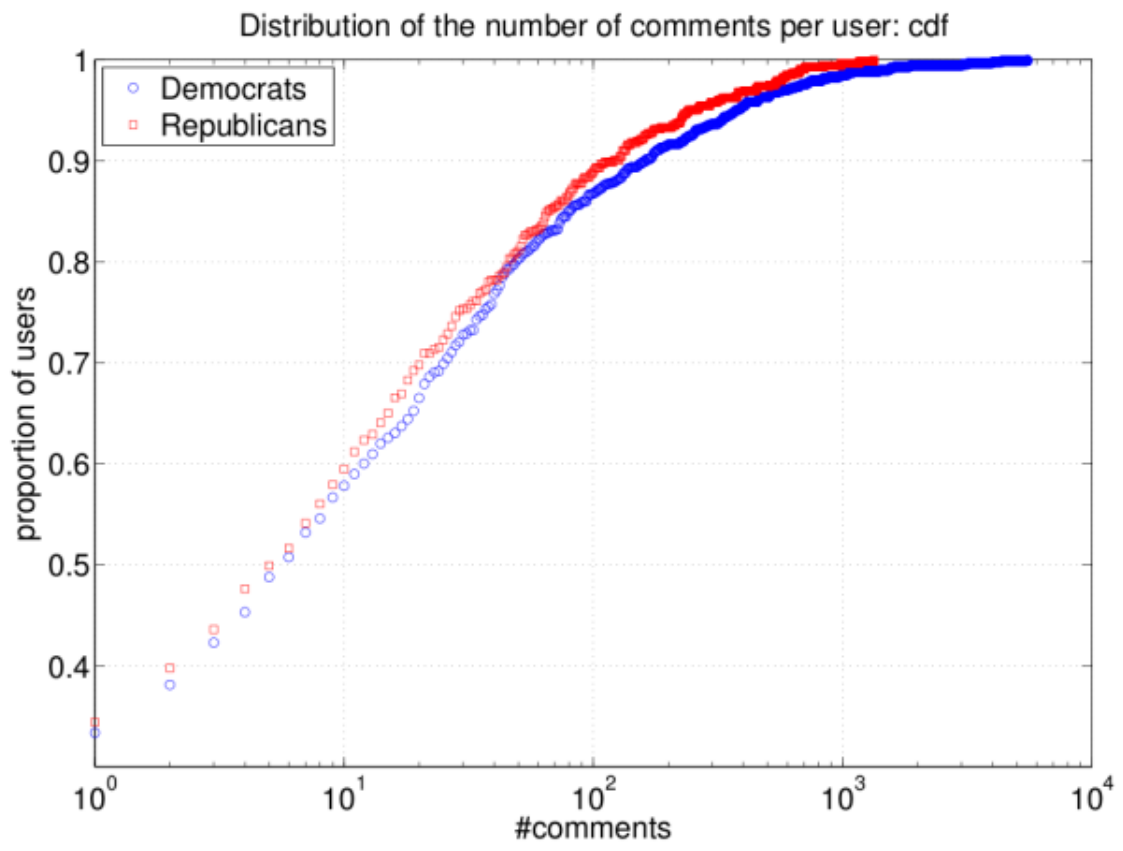

**Figure 3**: Cumulative distribution of the number of comments per user, for Democrats (blue) and Republicans (red).

comments, and 20% wrote more than 50. Again, we find the most active users among Democrats.

To analyze patterns of communication among our set of users, we identified two networks of interactions based on messages written by the users in talk pages. From article talk pages we extracted a *reply network*, by establishing a connection from user A to user B if user A has replied to a comment written by user B. From *user walls* we extracted a *wall network* by connecting user A to user B if user A has left a message on user B's personal talk page. Basic statistics of the two resulting networks are given in Table 1. It should be noted that the networks do not include all users in our sample, but only the ones who have interacted with other users in the sample. As a result, the reply network includes only 270 users (161 Democrats and 109 Republicans), while the wall network includes 435 users (258 Democrats and 176 Republicans).

**Table 1**: Basics statistics of the two interaction networks.

| | Nodes | Dem | Rep | Edges | Giant comp. | Reciprocity | Clustering coefficient | Average distance |
|---|---|---|---|---|---|---|---|---|
| Reply | 270 | 161 | 109 | 430 | 83.9% | 0.28 | 0.040 | 4.75 |
| Wall | 434 | 258 | 176 | 997 | 95.6% | 0.29 | 0.074 | 4.00 |

**Most edited articles**

Lists of the most frequently edited articles among Democrats, Republicans, and all Wikipedia editors were generated. The number of articles shared by various groups was calculated. The articles were also coded according to topics. Articles were coded as "Political" (relating to a political issue or a politician, e.g. United States Presidential Election, 2008; George Bush) or "Not Political" (e.g. Britney Spears, 2008 Summer Olympics). Political pages further coded as either, "Conservative" (related to a conservative politician, commentator, or issue, e.g. Rush Limbaugh), "Liberal" (related to a liberal politician, commentator, or issue, e.g. Al Gore), or "Neutral" (political in nature, but not partisan, e.g. European Union, September 11 attacks).

**Cross-party interactions**

To assess whether there is a preference for interaction among editors belonging to the same party, or to different parties, we studied the mixing coefficient of the networks, and we performed a shuffle test in order to assess statistical significance.

We first extracted from each network a matrix representing how many connections (based on comments) are directed from a Democrat to a Democrat, from a Democrat to a Republican, and so on. We then normalized these matrices and computed the mixing coefficient as the preference for inter-party or for intra-party interaction, according to Newman (2003). To assess statistical significance, we contrasted the results with a sample of randomized equivalents of the networks. More specifically, keeping the users fixed, (both in terms of their party affiliations and their numbers of in-coming and out-going links), we randomized the links between them, thereby generating a sample of networks characterized by the same structural properties, where the sample of users is the same but they are connected differently. We computed the average value $r_{\text{rand}}$ of the mixing coefficient in these networks, and the standard deviation $\sigma_{rand}$; finally we computed the *Z-score* as the difference between the value observed in the real network and the average over the randomized networks (Z-score = $(r - r_{\text{rand}})/\sigma_{rand}$). High positive values of Z indicate a preference for inter-party interactions, while high negative values, a preference for intra-party interactions. Results low in absolute values ($|Z| < 2$) correspond to neutral mixing, i.e. no statistically significant preference for either inter- or intra-party interaction (Foster, 2010).

**Emotions by party**

To infer the emotional content of the comments in the discussions, we used the Affective Norms for English Language Words (ANEW), a list of 1034 words that were scored by human raters on a 9-point scale along three emotional dimensions: valence, arousal, and dominance (Bradley & Lang, 1999).

- Valence measures the extent to which people associate these words with happiness, satisfaction and hope. Scores of 9 indicate high levels of positive feelings, and scores of 1 indicate high levels of negative feelings.
- Arousal captures the association of words with feelings of excitement, anger or frenzy (high ANEW score) and their opposites.

- Dominance, in turn, focuses on feelings of domination or being in control (high ANEW score) versus feelings of submission or awe.

We followed the method proposed in Dodds and Danforth (2010) to extract emotional scores from the discussions. For the set of comments of a given user in article talk pages, we counted the number of occurrences of the different ANEW words and calculated weighted averages over each of the corresponding ANEW scores. We then calculated macro-averages of these quantities, i.e. the mean value over the user averages per party.

**Conflict**

We examined discussion thread conflict in order to assess whether or not users exhibit different interaction styles with same party members versus different party members. All discussion threads on article talk pages that included at least two Democrats (577), at least two Republicans (154), or at least one Democrat and one Republican (584) from our sample were extracted from the data set. From this sample, we then selected all threads related to articles that dealt with political or other potentially controversial topics. Examples include "War on Terrorism", "Mike Huckabee", "Eliot Spitzer", and "Mahmoud Ahmadinejad." These threads were then coded for whether or not they were conflictual, and for whether or not the conflict was political in nature. This resulted in 132 threads with two or more Democrats, 76 threads with two or more Republicans, and 139 threads with at least one Democrat and one Republican. Holsti's (1969) reliability score was used to assess intercoder reliability. At .82, the coefficient was acceptable.

# Results

## Identity analysis

First, we tested to see if there were differences in the average number of userboxes listed on the profiles of Republicans and Democrats. There was no significant difference ($p = 0.3$). Unsurprisingly, Republicans ($M = 3.06$, $SD = 5.4$) had a significantly higher number of conservatively valenced user boxes than Democrats ($M = .08$, $SD = .44$) ($t = 4$, $p < .001$), while Democrats ($M = 2.51$, $SD = 3.47$) had a significantly higher number of liberally valenced user boxes than Republicans ($M = .27$, $SD = .60$) ($t = 4$, $p < .001$). Next we looked to see if there were any differences in number of “Wikipedia” listed on the user pages of Democrats and Republicans, but we did not find any ($p = 0.07$). Finally, we examined differences between the number of “Wikipedia” boxes and “Party” boxes for members of each party. Democrats had significantly more “Wikipedia” boxes ($M = 4.7$, $SD = 7.01$) than “Party” boxes ($M = 1.44$ , $SD = 1.34$), $t = 3.1$, $p < .01$. Republicans also had significantly more “Wikipedia” boxes ($M = 3.16$, $SD = 4.00$) than “Party” boxes ($M = 1.26$, $SD = .92$), $t = 3.5$, $p < .001$. Table 2 provides descriptive statistics for the identity analysis.

**Table 2**: Average Number of Userboxes per user.

| | Total nº of Boxes | Political Boxes | | | Party Boxes | Wikipedia Boxes |
|---|---|---|---|---|---|---|
| | | Conservative | Liberal | Other | | |
| Democrats | 49.24 | .08 | 2.51 | 2.20 | 1.44 | 4.7 |
| Republicans | 42.90 | 3.06 | .27 | 2.52 | 1.26 | 3.16 |

**Most edited articles**

Out of the 100 most edited articles, Democrats and Republicans had 44 articles in common. For Democrats, 38 of the top 100 most edited articles dealt with political topics. Out of those, 15 were coded as liberal, 15 were coded as conservative, and 8 were coded as neutral. Thirty-five out of the top 100 most edited articles by Republicans dealt with political topics. Of these, 7 were coded as liberal, 17 were coded as conservative, and 11 were coded as neutral. These findings stand in contrast to the most edited article for users in general, only 22 of which dealt with political topics. Of those, 3 were coded as conservative, 5 were coded as liberal, and 14 were coded as neutral. Table 3 contains an overview of the top 10 most edited articles by Democrats, Republicans, and Wikipedians in general.

**Table 3**: Top 10 Articles per number of distinct editors among Democrats, Republicans, and all users. Articles related to U.S. politics are indicated in bold.

| | Democrats | Republicans | All Users |
|---|---|---|---|
| 1. | **Barack Obama** | **George W. Bush** | **George W. Bush** |
| 2. | **Unites States presidential election, 2008** | **Unites States presidential election, 2008** | Wikipedia |
| 3. | **George W. Bush** | United States | United States |
| 4. | Unites States | **Republican Party (United States)** | **Barack Obama** |
| 5. | **Bill Clinton** | **John McCain** | Adolf Hitler |
| 6. | **Democratic party (United States)** | **Barack Obama** | Michael Jackson |
| 7. | Wikipedia | Wikipedia | Britney Spears |
| 8. | Britney Spears | **Ronald Reagan** | Jesus |
| 9. | **Hillary Rodham Clinton** | Virginia Tech Massacre | World War II |
| 10. | **Al Gore** | Adolf Hitler | PlayStation 3 |

**Cross-party interactions**

Tables 4 and 5 show the numbers of edges in the two networks under examination, broken down by party. Although the number of interactions between Democrats in Table 4 seems to be much larger at first sight, this is caused by the larger absolute number of Democrats in the network (see Table 1). This becomes visible in the results of a shuffle test for assortativity, shown in Table 6, which indicate that users exhibit no significant preference either for interacting with same party or different party users on article talk pages ($|Z| < 2$). That is, Democrats are not significantly more likely to interact either with other Democrats or with Republicans, nor are Republicans significantly more likely to interact either with other Republicans or with Democrats in the context of discussions on article talk pages. Figure 4 helps to understand this result, showing intra-party connections in blue (between Democrats) and red (between Republicans), and inter-party connections in yellow. The number of inter-party connections is remarkable.

An examination of the interactions on user walls shows a different pattern, indicating a significant preference for interaction among members of the same party (Z-score=3.33, see bottom row of Table 6) in this more personal communication space.

**Table 4**: Pairs of users interacting in article discussions, broken by party

| | Democrats | Republicans |
|---|---|---|
| Democrats | 193 | 94 |
| Republicans | 86 | 57 |

**Table 5**: Pair of users interacting on personal walls, broken by party

| | Democrats | Republicans |
|---|---|---|
| Democrats | 395 | 243 |
| Republicans | 187 | 172 |

**Table 6**: Mixing by party in the two interaction networks. r represents the mixing coefficient in the real network; $r_{rand}$ (avg) and $\sigma_{rand}$ the average and the standard deviation of the mixing coefficient over the randomised networks; Z-score the standard score. Values indicating statistically significant results (Z>2) are written in bold.

| | r | $r_{rand}$ (avg) | $\sigma_{rand}$ | Z-score |
|---|---|---|---|---|
| Reply | 0.070 | 0.0028 | 0.0505 | 1.33 |
| Wall | 0.095 | -0.0053 | 0.0301 | **3.33** |

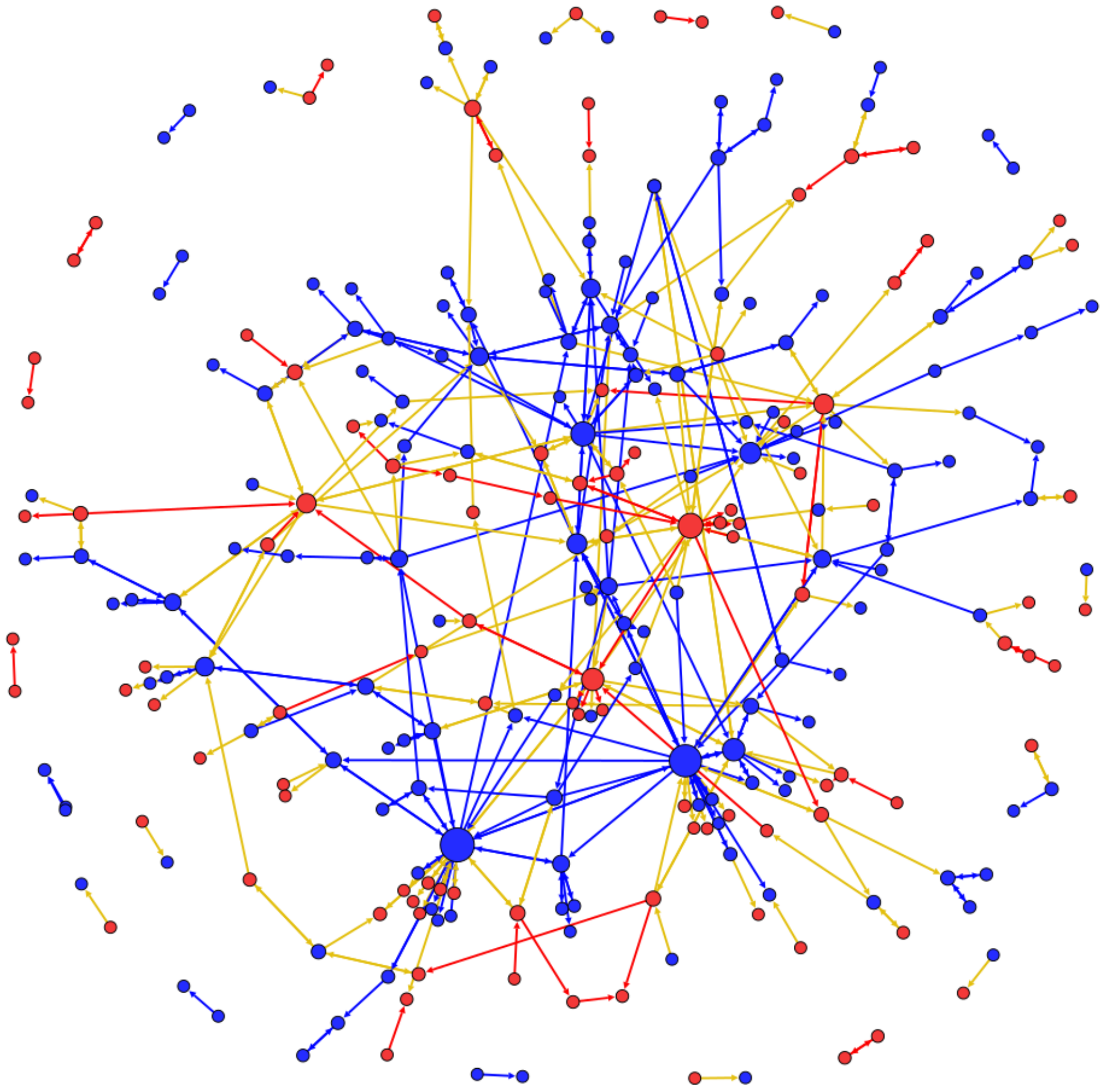

**Figure 4**: Reply network. Blue nodes represent Democrat users, and red nodes Republicans. The size of each node is proportional to the number of connections (degree). Edges connecting two Democrats are depicted in blue, edges connecting two Republicans in red, edges connecting a Democrat and a Republican in yellow.

**Emotions by party**

We conducted a sentiment analysis of the comments written in article talk pages and compared the scores of Democrats and Republicans. We restricted the analysis to the more active users whose comments contain in total at least 100 ANEW words (164 Democrats and 95 Republicans). The rationale for this restriction is that results of the analysis are more reliable with a larger sample of words per user, as they are not sensitive to single words written by less active editors. Results of this analysis indicate that Democrats write comments whose valence is significantly more positive (6.05 vs. 5.94, $p < 0.01$), and whose dominance is significantly greater (5.41 vs. 5.35, $p < 0.05$), than comments made by Republicans.

One could now wonder whether this difference is due to comments by Democrats being more positive than average, or to comments by Republicans being less positive than average. If we compare these values with the ones reported by Laniado, Castillo, Kaltenbrunner & Fuster-Morell (2012) for the same dataset considering all active users (with at least 100 comments), we find that the average valence of comments lies more or less in the middle (5.98), indicating that comments written by members of both parties deviate from the average, in two opposite directions. The overall average dominance (5.40) reported in Laniado et al. (2012) also lies between the values we observed for the two parties, but much closer to the average among Democrats.

The difference that we observed, both for valence and dominance, is not statistically significant if we normalize the score of each comment by the average of the entire discussion in which it is written. This suggests that the difference is mostly due to the discussions in which users participate and their topics, with Democrats engaged in discussions characterized by more positive emotions, and Republicans in discussions with a more negative tone and evoking feelings of being less in control.

To better illustrate this result, we show in Figure 5 the words belonging to the ANEW lexicon that are used more frequently by Republicans (in red) and Democrats (in blue). For example, the word 'war' is used more frequently in comments authored by Republicans than in comments by Democrats. Font size is proportional to the relative difference in frequency. Consistent with the results presented above, we only considered users with at least 100 ANEW words in their comments. Frequency of each word was first computed for each single user, and then averaged over parties.

Observing the tag cloud, we notice that the words 'war' and 'hurricane' are used more frequently by Republicans. Given the tendency of users to be more active on articles related to their party (as observed in Table 3), this may be related to the occurrence of hurricane Katrina and the wars in Afghanistan and Iraq during the presidency of George W. Bush. Figure 5 also helps to explain the results of the sentiment analysis based on the ANEW lexicon, with the Republican tag cloud showing terms such as 'war', 'church' or 'shadow', evoking more negative emotions or feelings of submission and awe. This resonates with the theory of Lakoff (2009), which links conservative discourse to concepts like "order based on fear", "intimidation and obedience", "authoritarian hierarchy". The Democratic tag cloud, on the other hand, with words such as 'color', 'city', or 'thought' shows more inclusive values such as "equality" and "social responsibility", concepts rather linked to progressive thinking (Lakoff, 2009, p. 1).

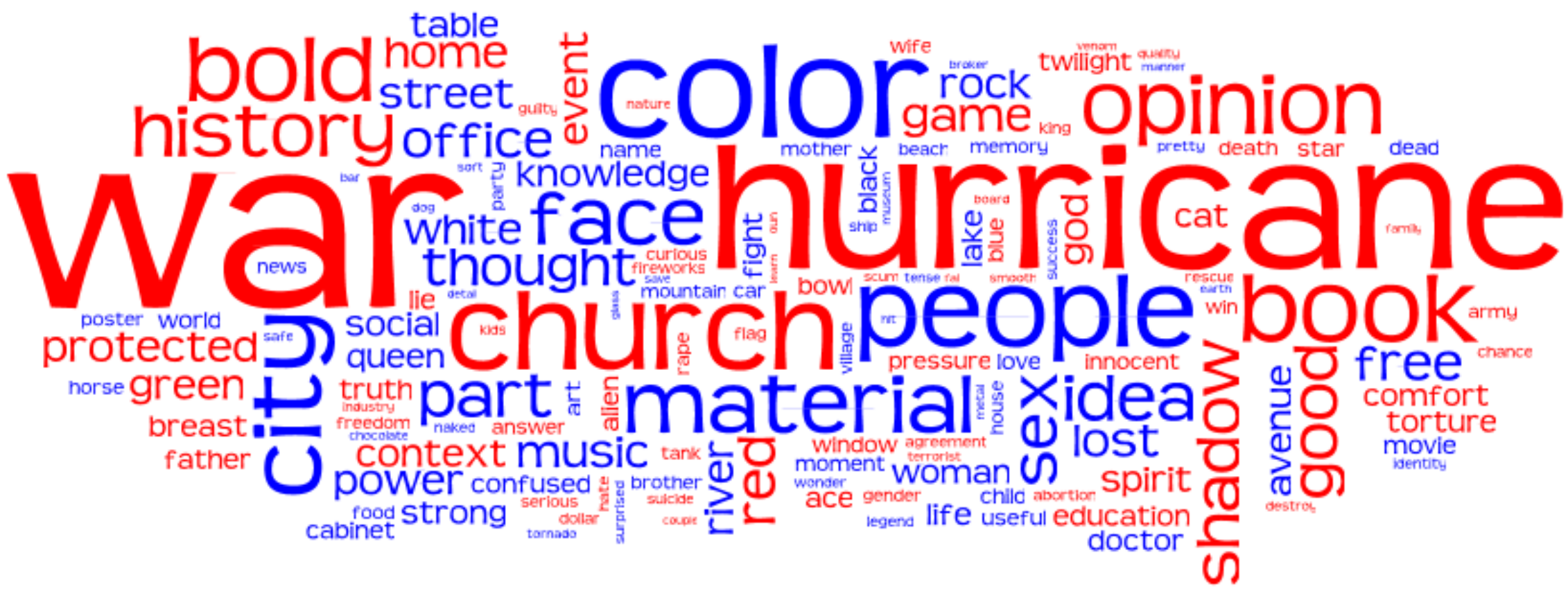

**Figure 5**: Tag cloud of words used more frequently by Democrats (blue) and by Republicans (red). Font size is proportional to the relative difference in frequency.

**Conflict analysis**

Finally, we examined levels of conflict in discussion threads that dealt with political or other potentially controversial topics. There were relatively high levels of conflict across all three groups of threads that we examined. This is not a particularly surprising finding, given that we purposefully selected threads that dealt with controversial topics. Sixty-six percent of the threads that contained at least two Democrats, 77% of threads that contained at least two Republicans, and 74% of threads that contained at least one Democrat and one Republican were coded as conflictive. See Table 7 for an overview. In order to assess differences in levels of conflict across the three groups, we use a Chi-square goodness-of-fit, with the null hypothesis that the proportion of conflict is the same for the three subsets as in totality. There was a significantly ($p < 0.05$) lower volume of conflict in the threads involving two Democrats, but there was no significant difference in the volume of conflict in the Republican ($p = 0.25$) and cross-party threads ($p = 0.23$).

**Table 7**: Number of politically valenced threads displaying conflict.

| | Involving a Democrat and a Republican | Involving two Democrats | Involving two Republicans |
|---|---|---|---|
| Total number of threads | 583 | 576 | 153 |
| Political threads | 144 | 134 | 77 |
| Political threads displaying conflict | 106 | 84 | 59 |

## Discussion

Our results paint an interesting, and somewhat mixed picture, of the nature of interactions among members of the Wikipedia community who espouse a political affiliation. First, we examined identity representation practices. We found that a subset of users on Wikipedia publicly proclaim their political affiliation through userboxes, and users who proclaim their affiliation for a particular party tend to have high numbers of userboxes that are ideologically aligned with that party. However, these 'political' users also had equally high numbers of Wikipedia userboxes. That is, boxes that espoused an identity of being a 'Wikipedian.' The results indicate that the social identities of being a member of a political party and being Wikipedian may be equally important. Analysis of patterns of activity and interaction indicates that which identity is activated may depend on context and the nature of activities in which users are engaged.

An examination of the most edited articles for each group reveals that Democrats and Republicans both exhibit a tendency for editing articles that deal with political topics. For both groups, roughly one-third of the most edited articles dealt with political topics, compared to less than one-quarter for users in general. Interestingly, for both groups we find a preference for topics directly related to their party, such as "Barack Obama" or "Bill Clinton" for Democrats, "John McCain" or "Republican Party" for Republicans.

Despite the preference for working on articles related to one's own party, when analyzing patterns of interaction in discussions about the encyclopedic content (i.e. in article talk pages), we found a neutral mixing pattern, indicating no preference for intra-party interactions. We also do not observe a preference for inter-party interactions, as we might expect if there were a prevalence of partisan discussions, with most users acting as "fighters" (Kelly, Fisher, & Smith, 2006) and engaging in disputes with users supporting the other party. Instead, we observe no significantly prevalent mixing pattern: when dealing with encyclopedic content, editors appear to be equally likely to engage conversations with users from the other party as with users from the same party.

In contrast, we did see evidence for preference to interact with members of the same party in user walls. It is interesting that we observe this tendency in these more personal spaces, but not on article talk pages. It may be that in the course of conducting activities that are central to the Wikipedia community (e.g. editing articles), the identity of being a Wikipedian is activated and, as a result, the political identity is not salient. In the context of interactions on user walls, where personal activities take greater precedence, the importance of political ideology may shine through more strongly.

Results of the sentiment analysis revealed that Democrats tend to write comments that are more positive and dominant than comments written by Republicans. However, we no longer see this tendency when we normalize by the average scores of the entire discussion. This is an interesting finding because it suggests that what we are observing is that, while Democrats may not be more positive or feel more in control than Republicans in general, they do seem to be involved in discussions that evoke more positive emotions and fewer sentiments of submission and awe.

Finally, we found that levels of conflict were high both within and across parties when the discussion threads dealt with political or other potentially controversial topics.

Interestingly, there were a significantly greater number of conflictive cross-party and Republican threads, indicating that Democrats have lower rates of within party conflict in the context of these controversial threads.

## Conclusions

Although Democrats and Republicans seem to maintain their political identity (as emerging from the table of most edited articles, and the tag cloud of words used more frequently in discussions), our findings show that users displayed more “Wikipedia” boxes than political boxes on their user pages, indicating that the identity of being a Wikipedian may be more salient in the context of this community. Further, the lack of preference to interact with same-party members in the context of article discussions does not indicate the same polarization that was observed in other contexts (Adamic & Glance, 2005; Conover et al., 2011). In this sense, the Wikipedian identity seems to predominate over party identity. Hence, the results of our analysis show that despite the increasing political division of the U.S., there are still areas in which political dialogue is possible and happens.